\documentclass[twocolumn,preprintnumbers,amsmath,amssymb]{revtex4}
\usepackage[dvips]{graphicx}
\usepackage{graphics}
\usepackage{amssymb}

\begin{document}

\title{Existence of A Rigorous Density-Functional Theory for Open Electronic Systems}
\author{Xiao Zheng}
\author{Fan Wang}
\author{GuanHua Chen}
\email{ghc@everest.hku.hk} \affiliation{Department of Chemistry,
The University of Hong Kong, Hong Kong, China}

\date{\today}


\begin{abstract}
We prove that the electron density function of a real physical
system can be uniquely determined by its values on any finite
subsystem. This establishes the existence of a rigorous
density-functional theory for any open electronic system. By
introducing a new density functional for dissipative interactions
between the reduced system and its environment, we subsequently
develop a time-dependent density-functional theory which depends in
principle only on the electron density of the reduced system.
\end{abstract}

\maketitle

Density-functional theory (DFT) has been widely used as a research
tool in condensed matter physics, chemistry, materials science,
and nanoscience. The Hohenberg-Kohn theorem~\cite{hk} lays the
foundation of DFT. The Kohn-Sham formalism~\cite{ks} provides a
practical solution to calculate the ground state properties of
electronic systems. Runge and Gross extended DFT further to
calculate the time-dependent properties and hence the excited
state properties of any electronic systems~\cite{tddft}. The
accuracy of DFT or time-dependent DFT (TDDFT) is determined by the
exchange-correlation functional. If the exact exchange-correlation
functional were known, the Kohn-Sham formalism would have provided
the exact ground state properties, and the Runge-Gross extension,
TDDFT, would have yielded the exact time-dependent and excited
states properties. Despite their wide range of applications, DFT
and TDDFT have been mostly limited to isolated systems.

Many systems of current research interest are open systems. A
molecular electronic device is one such system. DFT-based
simulations have been carried out on such devices~\cite{prllang,
prlheurich, jcpluo, langprb, prbguo, jacsywt, jacsgoddard,
transiesta, jcpratner}. These simulations focus on steady-state
currents under bias voltages. Two types of approaches have been
adopted. One is the Lippmann-Schwinger formalism by Lang and
coworkers~\cite{langprb}. The other is the first-principles
nonequilibrium Green's function (NEGF) technique~\cite{prbguo,
jacsywt, jacsgoddard, transiesta, jcpratner}. In both approaches the
Kohn-Sham Fock operator is taken as the effective single-electron
model Hamiltonian, and the transmission coefficients are calculated
within the noninteracting electron model. The investigated systems
are not in their ground states, and applying ground state DFT
formalism for such systems is only an approximation~\cite{cpdatta}.
DFT formalisms adapted for current-carrying systems have also been
proposed recently, such as Kosov's Kohn-Sham equations with direct
current~\cite{jcpkosov} and Burke \emph{et al.}'s Kohn-Sham master
equation including dissipation to phonons~\cite{prlburke}. However,
practical implementation of these formalisms requires the electron
density function of the entire system. In this paper, we present a
rigorous DFT formalism for any open electronic system. The
first-principles formalism depends in principle only on the electron
density function of the reduced system, and can be used to simulate
the transient electronic response.

It is often implicitly assumed that the electron density function of
any real physical system is real analytic~\cite{ourarxiv}. This is
reflected in many existing first-principles methodologies. In
practical quantum mechanical simulations, real analytic functions
such as Gaussian functions, Slater functions and plane wave
functions are adopted as basis sets, which results in real analytic
electron density distribution. However, analyticity is not
guaranteed for all presently used basis functions. For instance, in
the linearized augmented plane wave (LAPW) method~\cite{booklapw},
radial functions times spherical harmonics span the inside of
muffin-tin spheres whereas plane waves are used for the interstitial
regime. By this approximation the electron density cannot be
analytically continued across the interface of the two disconnected
parts. It is thus interesting to ask whether the electron density
function of any real system is in principle real analytic. This
question was settled recently for time-independent systems.

Fournais~\emph{et al.} proved that the electron densities of atomic
and molecular eigenfunctions are real analytic in $\mathbf{r}$-space
away from the nuclei~\cite{analyticity}. Their proof is based on the
fact that the time-independent Schr\"{o}dinger equation,
$\hat{H}\Psi=E\Psi$, is an elliptic partial differential equation,
and its eigenfunctions can always be real analytic (except isolated
points) and quadratically integrable. Their rigorous proof implies
that the electron density function of a real physical system is real
analytic (except at nuclei) when the system in its ground state, any
of its excited eigenstates, or any state which is a linear
combination of finite number of its eigenstates. In their
derivation~\cite{analyticity}, nuclei are treated as point charges,
and this leads to non-analytic electron density at these isolated
points. Note that the isolated points at nuclei can be excluded for
the moment from the physical space that we consider, so long as the
space is connected. Later we will come back to these isolated
points, and show that their inclusion does not alter our results. We
show now that for a time-independent real physical system the
electron density distribution function in a sub-space determines
uniquely its values on the entire physical space. This is nothing
but analytic continuation of a real analytic function. The proof for
univariable real analytical functions can be found in textbooks, for
instance, Ref.~\cite{proof1}. The extension to multivariable real
analytical functions is straightforward.

{\it Lemma :} The electron density function $\rho(\mathbf{r})$ is
real analytic in a connected physical space $U$. $D\subseteq U$ is a
sub-space. If $\rho(\mathbf{r})$ is known for all $\mathbf{r}\in D$,
$\rho(\mathbf{r})$ can be uniquely determined on the entire space
$U$.

{\it Proof:} To facilitate our discussion, the following notations
are introduced. Set $\mathbb{Z}^{+} = \{0,1,2,\ldots\}$, and
$\gamma$ is an element of $(\mathbb{Z}^{+})^{3}$, \emph{i.e.},
$\gamma = (\gamma_{1},\gamma_{2},\gamma_{3})\in(\mathbb{Z}^{+})
^{3}$. The displacement vector $\mathbf{r}$ is denoted by the
three-dimensional variable $x = (x_{1},x_{2},x_{3})\in U$. Denote
that $\gamma\,\mbox{!} = \gamma_{1}\,\mbox{!}\:\gamma_{2}\,
\mbox{!}\:\gamma_{3}\,\mbox{!}\,$, $x^{\gamma} =
x_{1}^{\gamma_{1}}\:x_{2}^{\gamma_{2}}\:x_{3}^{\gamma_{3}}$, and $
\frac{\partial^{\gamma}}{\partial x^{\gamma}} =
\frac{{\partial}^{\gamma_{1}}}{\partial x_{1}^{\gamma_{1}}}
\frac{\partial^{\gamma_{2}}}{\partial x_{2}^{\gamma_{2}}}
\frac{\partial^{\gamma_{3}}}{\partial x_{3}^{\gamma_{3}}}$.

Suppose that another density distribution function $\rho'(x)$ is
real analytic in $U$ and equal to $\rho(x)$ for all $x\in D$. We
have $\frac{\partial^{\gamma}\rho(x)}{\partial x^{\gamma}} =
\frac{\partial^{\gamma}\rho'(x)}{\partial x^{\gamma}}$ for all $x\in
D$ and $\gamma\in(\mathbb{Z}^{+})^{3}$. Taking a point $x_{0}\in D$
and at or infinitely close to the boundary of $D$, we may expand
$\rho(x)$ and $\rho(x')$ around $x_{0}$, \emph{i.e.},
$\rho(x)=\sum_{\gamma\in(\mathbb{Z}^{+})^{3}}\frac{1}{\gamma !}
\left.\frac{\partial^{\gamma}\rho(x)}{\partial
x^{\gamma}}\right\vert_{x_{0}} (x-x_{0})^{\gamma}$ and
$\rho'(x)=\sum_{\gamma\in (\mathbb{Z}^{+})^{3}}\frac{1}{\gamma !}
\left.\frac{\partial^{\gamma}\rho'(x)}{\partial
x^{\gamma}}\right\vert_{x_{0}} (x-x_{0})^{\gamma}$. Assuming that
the convergence radii for the Taylor expansions of $\rho(x)$ and
$\rho'(x)$ at $x_{0}$ are both larger than a positive finite real
number $b$, we have thus $\rho(x)=\rho'(x)$ for all $x\in
D_{b}(x_{0})=\left\{x:\left\vert x-x_{0}\right\vert <b \right\}$
since $\left.\frac{\partial^{\gamma}\rho(x)}{\partial
x^{\gamma}}\right\vert_{x_{0}} =
\left.\frac{\partial^{\gamma}\rho'(x)}{\partial x^{\gamma}}
\right\vert_{x_{0}}$. Therefore, the equality $\rho'(x)=\rho(x)$ has
been expanded beyond $D$ to include $D_{b}(x_{0})$. Since $U$ is
connected the above procedures can be repeated until
$\rho'(x)=\rho(x)$ for all $x\in U$. We have thus proven that $\rho$
can be uniquely determined in $U$ once it is known in $D$.

With the above \emph{Lemma} we are ready to prove the following
theorem:

{\it Theorem 1:} A connected time-independent real physical system
is in the ground state. The electron density function
$\rho(\mathbf{r})$ of any finite subsystem determines uniquely all
electronic properties of the entire system.

{\it Proof:} Assuming the physical spaces spanned by the subsystem
and the connected real physical system are $D$ and $U$,
respectively. $D$ is thus a sub-space of $U$, \emph{i.e.},
$D\subseteq U$. According to the above lemma, $\rho(\mathbf{r})$ in
$D$ determines uniquely its values in $U$, \emph{i.e.},
$\rho(\mathbf{r})$ of the subsystem determines uniquely
$\rho(\mathbf{r})$ of the entire system.

Inclusion of isolated points, lines or planes where
$\rho(\mathbf{r})$ is non-analytic into the connected physical space
does not violate the above conclusion, so long as $\rho(\mathbf{r})$
is continuous at these points, lines or planes. This can be shown
clearly by performing analytical continuation of $\rho(\mathbf{r})$
infinitesimally close to them. Therefore, $\rho(\mathbf{r})$ of any
finite subsystem determines uniquely $\rho(\mathbf{r})$ of the
entire physical system including nuclear sites. Hohenberg-Kohn
theorem~\cite{hk} states that the ground state electron density
distribution of any system determines uniquely all its electronic
properties. Hence we conclude that ground state $\rho(\mathbf{r})$
of any finite subsystem determines all the electronic properties of
the entire real physical system.

As for time-dependent systems, the issue is less clear. Although it
seems intuitive that the electron density function of any
time-dependent real physical system is real analytical (except
isolated points in space-time), it turns out quite difficult to
prove the analyticity rigorously. Fortunately we are able to
establish a one-to-one correspondence between the electron density
function of any finite subsystem and the external potential field
which is real analytical in both $t$-space and $\mathbf{r}$-space,
and thus circumvent the difficulty concerning the analyticity of
time-dependent electron density function. For time-dependent real
physical systems, we have the following theorem:

{\it Theorem 2:} The electron density function of a real physical
system at $t_0$, $\rho(\mathbf{r},t_0)$, is real analytic in
$\mathbf{r}$-space, and the corresponding wave function is
$\Phi(t_0)$. The system is subjected to a real analytic (in both
$t$-space and $\mathbf{r}$-space) external potential field
$v(\mathbf{r},t)$. Let $D$ be a finite $\mathbf{r}$-subspace. The
time-dependent electron density function on the subspace $D$,
$\rho(\mathbf{r},t)$ with $\mathbf{r}\in D$, has thus a one-to-one
correspondence with $v(\mathbf{r},t)$ and determines uniquely all
electronic properties of the entire time-dependent system.

{\it Proof:} Let $v(\mathbf{r},t)$ and $v'(\mathbf{r},t)$ be two
real analytical potentials in both $t$-space and $\mathbf{r}$-space
which differ by more than a constant at any time $t\geqslant t_0$,
and their corresponding electron density functions are
$\rho(\mathbf{r},t)$ and $\rho'(\mathbf{r},t)$, respectively.
Therefore, there exists a minimal nonnegative integer $k$ such that
the $k$-th order derivative differentiates these two potentials at
$t_0$:
\begin{equation}
   \left.\frac{\partial^k}{\partial t^k}\left[v(\mathbf{r},t) -
   v'(\mathbf{r},t)\right]\right\vert_{t=t_0} \neq \mbox{const}.
   \label{v-vp}
\end{equation}
Following exactly the Eqs.~(3)-(6) of Ref.~\cite{tddft}, we have
\begin{eqnarray}
   \left.\frac{\partial^{k+2}}{\partial t^{k+2}}\left[
   \rho(\mathbf{r},t)-\rho'(\mathbf{r},t)\right]\right\vert_{t=t_0}
   &=& -\nabla\cdot u(\mathbf{r}), \label{rho-rhop}
\end{eqnarray}
where
\begin{eqnarray}
   u(\mathbf{r}) &=& \rho(\mathbf{r},t_0)\,\nabla\!\left\{
   \left.\frac{\partial^k}{\partial t^k}\left[v(\mathbf{r},t) -
   v'(\mathbf{r},t)\right]\right\vert_{t=t_0}\right\}.
   \label{uofr}
\end{eqnarray}
Due to the analyticity of $\rho(\mathbf{r},t_0)$, $v(\mathbf{r},t)$
and $v'(\mathbf{r},t)$, $\nabla\cdot u(\mathbf{r})$ is also real
analytic in $\mathbf{r}$-space. It has been proven in
Ref.~\cite{tddft} that it is \emph{impossible} to have $\nabla\cdot
u(\mathbf{r})=0$ on the entire $\mathbf{r}$-space. Therefore it is
also impossible that $\nabla\cdot u(\mathbf{r})=0$ everywhere in $D$
(otherwise based on the \emph{Lemma} proven earlier, $\nabla\cdot
u(\mathbf{r})$ can always be analytically continued from the
particular finite $\mathbf{r}$-subspace where its values are $0$ to
the entire $\mathbf{r}$-space, and this would lead to $\nabla\cdot
u(\mathbf{r})= 0$ everywhere in the entire $\mathbf{r}$-space). We
have thus
\begin{eqnarray}
   \left.\frac{\partial^{k+2}}{\partial t^{k+2}}\left[
   \rho(\mathbf{r},t)-\rho'(\mathbf{r},t)\right]\right
   \vert_{t=t_0}&\neq& 0 \label{rhow-rho'w}
\end{eqnarray}
for $\mathbf{r}\in D$. This confirms the existence of a one-to-one
correspondence between $v(\mathbf{r},t)$ and $\rho(\mathbf{r},t)$
with $\mathbf{r}\in D$. $\rho(\mathbf{r},t)$ on the subspace $D$
thus determines uniquely all electronic properties of the entire
system. This completes the proof of {\it Theorem 2}.

Note that if $\Phi(t_0)$ is the ground state, any excited
eigenstate, or any state as a linear combination of finite number of
eigenstates of a time-independent Hamiltonian, the prerequisite
condition in {\it Theorem 2} that the electron density function
$\rho(\mathbf{r},t_0)$ be real analytic is automatically satisfied,
as proven in Ref.~\cite{analyticity}. As long as the electron
density function at $t=t_0$, $\rho(\mathbf{r},t_0)$, is real
analytic, it is guaranteed that $\rho(\mathbf{r},t)$ on the
subsystem $D$ determines all physical properties of the entire
system at any time $t$ if the external potential $v(\mathbf{r},t)$
is real analytic.

According to \emph{Theorem 1} and \emph{2}, the electron density
function of any subsystem determines all the electronic properties
of the entire time-independent or time-dependent physical system.
This proves in principle the existence of a rigorous DFT-type
formalism for open electronic systems. All one needs to know is the
electron density of the reduced system. The challenge that remains
is to develop a practical first-principles formalism.

\begin{figure}
\includegraphics[scale=0.45]{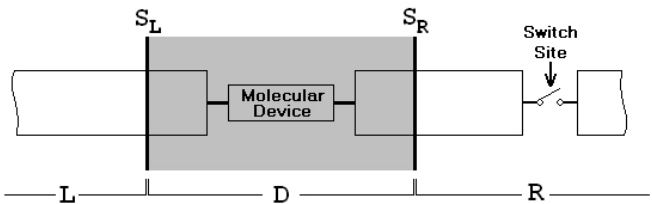}
\caption{\label{scheme} Schematic representation of the
experimental setup for quantum transport through a molecular
device.}
\end{figure}

Fig.~\ref{scheme} depicts an open electronic system. Region $D$
containing a molecular device is the reduced system of our
interests, and the electrodes $L$ and $R$ are the environment.
Altogether $D$, $L$ and $R$ form the entire system. Taking
Fig.~\ref{scheme} as an example, we develop a practical DFT
formalism for the open systems. Within the TDDFT formalism, a closed
equation of motion (EOM) has been derived for the reduced
single-electron density matrix $\sigma(t)$ of the entire
system~\cite{ldmtddft}:
\begin{equation}\label{eom4sigma0}
    i\dot{\sigma}(t) = [h(t),\sigma(t)],
\end{equation}
where $h(t)$ is the Kohn-Sham Fock matrix of the entire system, and
the square bracket on the right-hand side (RHS) denotes a
commutator. The matrix element of $\sigma$ is defined as
$\sigma_{ij}(t) = \langle a^{\dagger}_{j}(t)\,a_{i}(t)\rangle$,
where $a_{i}(t)$ and $a^{\dagger}_{j}(t)$ are the annihilation and
creation operators for atomic orbitals $i$ and $j$ at time $t$,
respectively. Fourier transformed into frequency domain while
considering linear response only, Eq.~(\ref{eom4sigma0}) leads to
the conventional Casida's equation~\cite{casida}. Expanded in the
atomic orbital basis set, the matrix $\sigma$ can be partitioned as:
\begin{equation}\label{matrixsigma}
    \sigma = \left[\begin{array}{lll}
    \sigma_{L}  & \sigma_{LD} & \sigma_{LR} \\
    \sigma_{DL} & \sigma_{D}  & \sigma_{DR} \\
    \sigma_{RL} & \sigma_{RD} & \sigma_{R}
    \end{array}\right],
\end{equation}
where $\sigma_{L}$, $\sigma_{R}$ and $\sigma_{D}$ represent the
diagonal blocks corresponding to the left lead $L$, the right lead
$R$ and the device region $D$, respectively; $\sigma_{LD}$ is the
off-diagonal block between $L$ and $D$; and $\sigma_{RD}$,
$\sigma_{LR}$, $\sigma_{DL}$, $\sigma_{DR}$ and $\sigma_{RL}$ are
similarly defined. The Kohn-Sham Fock matrix $h$ can be
partitioned in the same way with $\sigma$ replaced by $h$ in
Eq.~(\ref{matrixsigma}). Thus, the EOM for $\sigma_{D}$ can be
written as
\begin{eqnarray}\label{eom4sigmad0}
    i\dot{\sigma}_{D} &=& [h_{D},\sigma_{D}] + \sum_{\alpha=L,R}
    \left(h_{D\alpha}\sigma_{\alpha D}-\sigma_{D\alpha}
    h_{\alpha D}\right) \nonumber \\
    &=& [h_{D},\sigma_{D}] - i\sum_{\alpha=L,R}Q_{\alpha},
\end{eqnarray}
where $Q_{L}$ ($Q_{R}$) is the dissipative term due to $L$ ($R$).
With the reduced system $D$ and the leads $L/R$ spanned
respectively by atomic orbitals $\{l\}$ and single-electron states
$\{k_{\alpha}\}$, Eq.~(\ref{eom4sigmad0}) is equivalent to:
\begin{eqnarray}\label{eom4sigmad1}
    i\dot{\sigma}_{nm} &=& \sum_{l\in
    D}\,(h_{nl}\sigma_{lm}-\sigma_{nl}h_{lm}) - i\sum_{\alpha=L,R}
    Q_{\alpha,nm}, \label{eom4sigmad2} \\
    Q_{\alpha,nm} &=& i\sum_{k_{\alpha}\in\alpha}\big(h_{nk_{\alpha}}
    \sigma_{k_{\alpha}m}-\sigma_{nk_{\alpha}}
    h_{k_{\alpha}m}\big),\label{qterm0}
\end{eqnarray}
where $m$ and $n$ correspond to the atomic orbitals in region $D$;
$k_{\alpha}$ corresponds to an electronic state in the electrode
$\alpha$ ($\alpha = L$ or $R$). $h_{nk_{\alpha}}$ is the coupling
matrix element between the atomic orbital $n$ and the electronic
state $k_{\alpha}$. The current through the interfaces $S_L$ or
$S_R$ (see Fig.~\ref{scheme}) can be evaluated as follows,
\begin{eqnarray}\label{jcurrent}
    J_{\alpha}(t) &=& -\int_{\alpha}d\mathbf{r}\,\frac{\partial}
    {\partial t}\rho(\mathbf{r},t) \nonumber \\
    &=& -\sum_{k_{\alpha}\in\alpha}\frac{d}
    {dt}\,\sigma_{k_{\alpha}k_{\alpha}}\!(t) \nonumber \\
    &=& i\sum_{l\in D}\sum_{k_{\alpha}\in\alpha}\big(
    h_{k_{\alpha}l}\,\sigma_{lk_{\alpha}} -
    \sigma_{k_{\alpha}l}\,h_{lk_{\alpha}}\big) \nonumber \\
    &=& -\sum_{l\in D}Q_{\alpha,ll}
    = -\mbox{tr}\big[Q_{\alpha}(t)\big],
\end{eqnarray}
\emph{i.e.}, the trace of $Q_{\alpha}$.

At first glance Eq.~(\ref{eom4sigmad1}) is not self-closed since the
dissipative terms $Q_{\alpha}$ remain unsolved. According to {\it
Theorem 1} and \emph{2}, all physical quantities are explicit or
implicit functionals of the electron density of the reduced system
$D$, $\rho_{D}(\mathbf{r},t)$. Note that
$\rho_D(\mathbf{r},t)=\rho(\mathbf{r},t)$ for $\mathbf{r}\in D$.
$Q_{\alpha}$ is thus also a universal functional of
$\rho_{D}(\mathbf{r},t)$. Therefore, Eq.~(\ref{eom4sigmad1}) can be
recast into a formally closed form,
\begin{equation}\label{e4rdm}
  \!\!i\dot{\sigma_{D}}= \Big[h_{D}[\mathbf{r},t;\rho_{D}
    (\mathbf{r},t)],\sigma_{D}\Big]-i\!\!\sum_{\alpha=L,R}
    \!\!Q_{\alpha}[\mathbf{r},t;\rho_{D}(\mathbf{r},t)].
\end{equation}
Neglecting the second term on the RHS of Eq.~(\ref{e4rdm}) leads to
the conventional TDDFT formulation in terms of reduced
single-electron density matrix~\cite{ldmtddft} for the isolated
reduced system. The second term describes the dissipative processes
between $D$ and $L$ or $R$. Besides the exchange-correlation
functional, an additional universal density functional, the
dissipation functional $Q_{\alpha}[\mathbf{r},t;
\rho_{D}(\mathbf{r},t)]$, is introduced to account for the
dissipative interaction between the reduced system and its
environment. Eq.~(\ref{e4rdm}) is the TDDFT EOM for open electronic
systems. Burke \emph{et al.} extended TDDFT to include electronic
systems interacting with phonon baths~\cite{prlburke}, they proved
the existence of a one-to-one correspondence between
$v(\mathbf{r},t)$ and $\rho(\mathbf{r},t)$ under the condition that
the dissipative interactions (denoted by a superoperator
$\mathcal{C}$ in Ref.~\cite{prlburke}) between electrons and phonons
are fixed. In our case since the electrons can move in and out the
reduced system, the number of the electrons in the reduced system is
not conserved. In addition, the dissipative interactions can be
determined in principle by the electron density of the reduced
system. We do not need to stipulate that the dissipative
interactions with the environment are fixed as Burke \emph{et al.}.
And the only information we need is the electron density of the
reduced system. In the frozen DFT approach~\cite{warshel} an
additional exchange-correlation functional term was introduced to
account for the exchange-correlation interaction between the system
and the environment. This additional term is included in
$h_D[\mathbf{r},t;\rho_{D}(\mathbf{r},t)]$ of Eq.~(\ref{e4rdm}). An
explicit form of the dissipation functional $Q_{\alpha}$ is required
for practical implementation of Eq.~(\ref{e4rdm}). Admittedly
$Q_{\alpha}[\mathbf{r},t; \rho_{D}(\mathbf{r},t)]$ is an extremely
complex functional and difficult to evaluate. As various
approximated expressions have been adopted for the DFT
exchange-correlation functional in practical implementations, the
same strategy can be applied to the dissipation functional
$Q_{\alpha}$. Work along this direction will be published
elsewhere~\cite{longpaper}.

Given $Q_{\alpha}[\rho]$ how do we solve Eq.~(\ref{e4rdm}) in
practice? Again take the molecular device shown in Fig.~\ref{scheme}
as an example. We may integrate Eq.~(\ref{e4rdm}) directly by
satisfying the boundary conditions at $S_{L}$ and $S_{R}$. The only
boundary condition we need is the potentials at $S_{L}$ and $S_{R}$.
We need thus integrate Eq.~(\ref{e4rdm}) together with a Poisson
equation for Coulomb potential. And the Poisson equation is
subjected to the boundary condition determined by the potentials at
$S_{L}$ and $S_{R}$. It is important to point out that although in
principle its physical span can be small, in practice the reduced
system is to be chosen so that Eq.~(\ref{e4rdm}) can be solved
readily with convenient boundary conditions. For instance, for the
molecular electronic device depicted in Fig.~\ref{scheme}, the
reduced system $D$ contains not only the molecular device itself,
but also portions of the left and right electrodes. In this way the
Coulomb potential at the boundary take approximately the values of
the bulk leads.

To summarize, we have proved rigorously the existence of a
first-principles method for both time-independent and time-dependent
open electronic systems, and developed a formally closed TDDFT
formalism by introducing a new dissipation functional. This new
functional $Q_{\alpha}$ depends only on the electron density
function of the reduced system. With an explicit form for the
universal dissipation functional $Q_{\alpha}$, the time evolution of
an open electron system in external fields is fully characterized by
Eq.~(\ref{e4rdm}). In practical calculations, we need thus focus
only on the reduced system with appropriate boundary conditions.
This work greatly extends the realm of density-functional theory.

\begin{acknowledgements}
Authors would thank Hong Guo, Shubin Liu, Jiang-Hua Lu, Zhigang
Shuai, K. M. Tsang, Jian Wang, Arieh Warshel and Weitao Yang for
stimulating discussions. Support from the Hong Kong Research Grant
Council (HKU 7010/03P) is gratefully acknowledged.
\end{acknowledgements}

\end{document}